\documentclass[twocolumn,preprintnumbers,amsmath,amssymb,nofootinbib
,superscriptaddress]{revtex4}
\usepackage{hyperref}
\usepackage{graphicx}% Include figure files
\usepackage{color}
\usepackage{xcolor}

\newcommand{\be}{\begin{equation}}  
\newcommand{\ee}{\end{equation}}  
\newcommand{\bear}{\begin{eqnarray}}  
\newcommand{\eear}{\end{eqnarray}}  
\newcommand{\ba}{\begin{array}}  
\newcommand{\ea}{\end{array}}

\usepackage{diagbox}

\definecolor{rossoCP3}{cmyk}{0,.88,.77,.40}
\definecolor{blueRef}{rgb}{0.2,0.2,0.6}
\definecolor{blue}{rgb}{0,0.396,0.741}
\hypersetup{
	colorlinks, 
	bookmarksopen, 
	bookmarksnumbered,
	citecolor=blueRef, 		%color of links to bibliography
	linkcolor=rossoCP3,	%color of internal links
	urlcolor=rossoCP3,			%color of external links 
}

\newskip\humongous \humongous=0pt plus 1000pt minus 1000pt

\newif\ifdtup

\def\oldreffmt#1{\rlap{[#1]} \hbox to 2\parindent{}}

\def\figfmt#1{\rlap{Figure {#1}} \hbox to 1in{}}  
  
\def\etal{\hbox{\it et al.}}  
  
\def\beq{\begin{equation}}  
\def\eeq{\end{equation}}  
\def\bea{\begin{eqnarray}}  
\def\eea{\end{eqnarray}}

\def\bq{\begin{quote}}  
\def\eq{\end{quote}}

\def \lta {\mathrel{\vcenter  
     {\hbox{$<$}\nointerlineskip\hbox{$\sim$}}}}  
   
\def \etal {{\it et al.}\ }  
\relax

\newdimen\tdim  
\tdim=\unitlength  
\def\bar{\overline}

\begin{document}
\preprint{FERMILAB-PUB-20-086-T}

\title{ Composite Higgs Bosons and  Mini Black Holes }

\author{Christopher T. Hill}\email{hill@fnal.gov}
\affiliation{Theoretical Physics Department, \\
Fermi National Accelerator Laboratory, \\
P. O. Box 500, Batavia, IL 60510, USA}

\begin{abstract}

Pairs of standard model fermions can annihilate 
to produce mini black holes 
with gauge quantum numbers of the Higgs boson at $M_{Planck}$.  This
leads to a Nambu-Jona-Lasinio model at the Planck scale with strong coupling
which binds fermion pairs into Higgs fields. At critical coupling
the renormalization group dresses these objects, which then descend in scale
to emerge as bound-state Higgs bosons at low energies.  
We obtain the multi-Higgs spectrum 
of a ``scalar democracy.''  The observed
Higgs boson is a gravitationally bound $\bar{t}t$ composite;
sequential states are gravitationally bound composites of all SM fermion pairs,
where the lightest ones may be seen at the LHC and/or its upgrades.

\end{abstract}

\maketitle
 
\date{\today}

%\pacs{14.80.Bn,14.80.-j,14.80.-j,14.80.Da}

\section{Introduction}

Standard arguments suggest that a sufficiently energetic
collision between, {\it e.g.},  
a left-handed electron $(e_L)$ and an anti-right-handed electron $(\bar{e}_R )$,
%(which is left-handed) 
can produce a mini black hole $B$:
\bea
e_L+\bar{e}_R\rightarrow B
\eea
Production of $B$ requires $M_B=\sqrt{s}$ and that
the collision have an impact parameter,
$b$, where  $b \lta 2G\sqrt{s}$, hence
$b$ is the Schwarzschild
radius for the corresponding black hole.

Let us assume the total angular momentum  of the initial
state  is $s$-wave and spin zero. 
The incident charged electrons have the 
standard model
weak isospin, and hypercharge,  $[I_3, Y]$,
 $e_L\sim [-1/2, -1]$ and $\bar{e}_R\sim [0, 2]$.   
These will produce an electrically neutral black hole, $B\sim
[-1/2, 1]$, where the electric charge is 
$Q=I_3+\frac{Y}{2}= 0$. 
If we replace the incident $e_L$ by the
left-handed neutrino, $\nu_L\sim [1/2, -1]$,
we obtain the charged black hole. $B\sim [1/2, 1]$,
with $Q=1$. 

These are the quantum numbers of the neutral 
and charged components of the Higgs doublet in the standard model (SM).  
Therefore, SM fermions and gravity, alone, automatically 
imply scalar ``Higgs
bosons'' that are gravitationally bound pairs of
fermions, alas with masses  $\sim M_P$ !
Classically we describe these by the Reissner-Nordstrom (RN)
metric and  the Higgs black hole is an electroweak isodoublet,
has ``hair,''  with external gauge fields, $W^\pm$,
$Z^0$ and $\gamma$.  By conventional wisdom
they are guaranteed to exist. 

\subsection{A Black Hole --- Higgs Boson Connection?}

In the present paper we consider the possibility that
there is a deeper connection between the existence
of these ``Higgs black holes'' (HBH) 
and the physically observable Higgs boson(s) of an extended
standard model. 
While speculative, we think the correspondence of black hole
states to Higgs isodoubets is striking and may be 
a harbinger of a new physical reality.  This is a first
attempt to connect these, whereby the Higgs bosons emerge as composite
particles induced by the presence of mini-black holes. This is
therefore not a model in which we introduce a plethora of
new states, but rather a dynamical hypothesis that these
state are ``chemical''' and produced by existing degrees of
freedom at the gravitational scale.  

We directly consider the  virtual effects of the threshold HBH
and find that these may
imply a strong interaction as one approaches the Planck scale. 
We will take a Wilsonian approach and treat the interactions
of fermion pairs and mini-black holes
with an operator product expansion.  In essence, we are 
below the scale of the new black-hole states and their virtual
effects will induce new strong four-fermion interactions.

Integrating out the holes leads to an effective Nambu-Jona-Lasinio
model that then drives the formation of composite states. 
By renormalization
group (RG) effects, the black holes may
form composite Higgs fields in the infrared.  
One can view this as black holes
becoming dressed by the renormalization group
{\it i.e.}, becoming ``Wilsonian black holes'' as cores of Higgs bosons \cite{odint}.
However, from our point of view the physics is determined by the operator product expansion
below the threshold for mini black hole production.

The main issue is, how far into the infrared can this
composite spectrum extend?
At larger distance the composite Higgs bosons are mainly 
loops of SM particles and the HBH is virtual.
Fermion loops bind, subtracting from the 
bare mass of an HBH and pull it into the infrared. 

We would require an exact critical coupling of fermions to
the HBH to make the composite Higgs states 
 massless. This is analogous to criticality
in second order phase transitions.
However, in the present Nambu-Jona-Lasinio model
this involves a drastic fine-tuning (this is identical
to what happens in top-condensation models \cite{BHL,BH}). 

This drastic fine tuning, however is not new.  This
is the usual quadratic fine tuning that plagues the Higgs
boson of the standard model, and which thus plagues most theories of
electroweak symmetry breaking.  It remains unsolved.  Here we argue
that the fine tuning is
common to all fermion pair channels due to symmetry.
This implies the formation of
many scalars, one complex scalar per channel, and a universality
of the low energy couplings of fermions to the these boundstates.

We do not presently offer a solution
to the origin of the fine tuning problem, but a number of ideas come to mind and
we hope to develop them elsewhere.  The effective coupling of a fermion
pair to a composite scalar is dynamical, and may be treatable
in a variational calculation.
For critical coupling the fine tuning is a symmetry,
an effective scale invariance
of the bound-state with respect to the Planck mass.
Conceivably this might arise dynamically, 
{\it i.e.}, the bound-state system may internally
self-adjust dimensionless parameters to  minimize its energy,
and find the cancellation that realizes the symmetry.
This would likely be sensitive to the
quantum numbers of the composites, {\it e.g.}, favoring light color singlets
and leaving colored states at very large masses, and would dictate
the key features of the low energy spectrum of Higgs boundstates.
The nonzero resulting small masses for
the composite Higgs bosons would then arise from 
additional infrared scale breaking
effects, of order $10^2$ GeV to $\sim 10^6$ TeV.

Presently we'll assume something like this works, fine tune, and proceed.
We then
find that multiple Higgs scalars occur, at least one 
for any  $s$-wave fermion bilinear channel present at the Planck scale.
If all SM fermions are present near $M_P$ then we can form $1176$ 
complex scalars,
the symmetric bilinear representation of $SU(48)$.
This leads to $18$ Higgs doublets in the quark sector 
and $18$ in the lepton sector. This is an idea proposed
recently of ``scalar democracy'' \cite{HMTT,HMTT2,HP}. It 
is consistent with, and in principle ``explains,'' flavor physics.
It is
testable at the LHC upgrades
and it implies a plethora of new states for a $\sim 100$ TeV machine.

We are mainly interested in the physics as we approach the
threshold of a spectrum of black holes.  
Previous analyses of black hole production focus
on large $\sqrt{s}>>M_P$ in compactified extra dimensional schemes with
low effective $M_P$ in a string theory. These do not address
the threshold behavior and are not useful to us.
For the quantum theory near threshold we expect
a breakdown of classical intuition, just as is the case of the Hydrogen
atom. Here we find the ideas of Dvali and Gomez (DG) \etal
to be compelling and yield a useful ``portrait'' of the threshold theory
\cite{Dvali1,Dvali2,Dvali3} (see also \cite{Calmet}).

\subsection{Dvali-Gomez ``Portrait'' of a Mini Black Hole}

We briefly summarize the ideas of Dvali and Gomez (DG).
Here  black holes are composed
of ``condensates''  of a large number, $N$, of  gravitons and perhaps other objects
such as the fermion pair that creates an HBH.
The Dvali-Gomez theory is intrinsically a strongly coupled
gravity as one approaches the Planck scale. 
The  behavior becomes classical as $N>>1$ and we would expect
the geometrical aspects of black holes are then emergent.

On the other
hand, for small $N\rightarrow 1$  we approach the quantum limit, 
and the behavior is radically different than the classical
picture. Here 
many classical theorems about black holes break down
(such as the ``no-hair'' theorem; moreover the viabilty of
global symmetries, such as flavor symmetries is maintained). 
For small $N$ the states have quantized masses (modulo widths)
and form a tower of resonances with schematic
decay chains that cause transitions $N\rightarrow N-1$ (Hawking radiation).
The RN-black hole ``remembers'' the global charges that produced it.
Near threshold, the decay width of small $N$ black holes approaches $\sim M_P$. 
The effective coupling of matter to threshold black holes is strong.

A threshold Schwarzschild black hole consists of a single graviton 
with mass $\mu\sim \pi/2R$, localized within the  Schwarzschild radius $R$.
The graviton can be thought of as a half-wave ``lump'' within the (effective) horizon 
of size $\sim 2R$, and corresponding to
a full wavelength of $\sim 4R$. 
 If we 
consider a Fock state with $N$ quanta in this mode, we
will have a black hole mass $M=N\mu =N\pi/2 R$, which  will form a
horizon as:
\bea
1=2GM/R =GN\pi/R^2, \;\makebox{hence}\;  R=\sqrt{\pi N}M_P^{-1},
\eea
where $G=1/M_P^2$, 
and therefore:
\bea
M_N=N\pi/2 R=\sqrt{N\pi}M_P/2.
\eea

A key feature of
the DG theory is that it has an effective smallest quantum
wavelength
and corresponding momentum cutoff.
For concreteness, we will define these to be, respectively:\footnote{Note, we have inserted the 
necessary factors of $\pi$ into the DG discussion
when one relates ``wavelength'' or ``Schwarzschild radius'' 
to ``mass'' or ``momentum''
when one sets $\hbar=1$.}
\bea
\lambda_0\sim \sqrt{\pi}M_P^{-1}, \qquad  p_0 \sim 2\pi /\lambda_0=\sqrt{4\pi}M_P.
\eea
We've defined 
$\lambda_0$  as the Schwarzschild radius of the single graviton
black hole in the DB picture, $N=1$.
At shorter distances the gravitational
interaction is so strong that ordinary space-time becomes unthinkable.
Anything with a quantum wavelength $\lta \lambda_0$ will be self-cloaked
in gravitons, {\it e.g.}, if one imagines boosting an electron
above the cutoff momentum, say to $\sim 2p_0$, one will have a
pointlike electron with momentum  $\sim p_0$ and collinear gravitons
with $\sim p_0$. 
Hence at
short distances we can never resolve a pointlike electron with
momentum component in excess of the cut-off.

Therefore, the smallest threshold black hole has 
a Schwarzschild radius $R=\sqrt{\pi}M_P^{-1}$, and a constituent
quantum wavelength $\lambda=4\sqrt{\pi}M_P^{-1}$, safely
larger than the fundamental wavelength cut-off 
$\lambda > \lambda_0$.
As $N$ increases, the black hole size does as well, $\propto \sqrt{N}$. Higher
modes then become accessible, never exceeding  the fundamental cutoff
momentum $p_0$. 

It is important to keep in mind that $N$ is the occupancy
of a mode, and not a ``principle quantum number'' of the modes.
DG refer to large $N$ as a ``Bose-Einstein condensate;'' these
are actually Fock states, until the black hole Schwarzschild radius gets large and
more available modes with wavelength greater than the cut-off open up.
As we excite a black hole its radius grows as $\sqrt{N}$ and
we produce more gravitons in the lowest mode, and  
the wavelengths of these constituents is never smaller than
$\lambda_0$. Conversely, we see that $N\propto R^2$ which is
an affirmation of Bekenstein entropy in the classical limit.
This is also
the basis of the claim of DG that Einstein gravity is self-healing
and ``classicalizes'' in the far UV, and does not require a UV completion
theory.

\section{Mini Black Hole Induced Higgs Compositeness}

We can extend the DG model to Reissner-Nordstrom black holes
by including the incident fermions as components of the black hole. 
The ground-state then consists of
the pair of incident fermions that produced it,
${f}_1f_2$. The $N$th excitation (occupancy) above the ground-state
will have these two fermions plus $N$ gravitons. Each is assumed to have an 
energy $\mu\sim \hbar\pi /2R$ where $R$ is the Schwarzschild radius. 
The system 
self-binds into a black hole with mass $M_N=(2+N)\pi/2R_N$.
Hence:
\bea
\frac{G_N(2+N)\pi}{R^2_N}&=&1 \qquad R_N=\frac{\sqrt{(2+N)\pi}}{M_P}
\nonumber \\
M_N&=& {\sqrt{(2+N)\pi}M_P/2}.
\eea
%In a sense, for small $N$ this is a ``bag model'' of a black hole.\footnote{The term
%``black hole bag model'' was coined by Bill Bardeen in discussions.}  
To expedite the discussion we focus  on  
a single flavor channel, and only the ground-state black hole 
of mass $M_0= ({\sqrt{\pi/2}}){M_P}$ and
Schwarzschild radius $R_0={\sqrt{2\pi}}/{M_P}$.

%\newpage

\subsection{Effective Field Theory}

We presently assume that the incident flavors
are a pair  consisting of the
electron doublet
 $E_L = (\nu,e)_L$  and anti-right-handed 
 singlet $\bar{e}_R$.   
Therefore the produced RN black hole, $B_0$, will be an HBH weak isodoublet
with quantum numbers of the SM Higgs doublet $B_0 \sim \bar{e}_{R}E_L $.

Consider an effective  field theory
of the coupling of the leptons to the
threshold HBH $B_0$:
\bea
\label{L0}
{\cal{L}}_{0} &=&
DB_0^\dagger DB_0-g\left(\bar{E}_{L}B_0 e_R + h.c.\right) 
\nonumber \\
&& -M_0^2 B_0^\dagger B_0.
\eea
While this is a local approximation, which cannot
be an exact description of the production process,
the purpose of this effective field theory is only to roughly estimate the
coupling  constant $g$.

We compute the field theory cross-section for 
$E_L +\bar{e}_{R}\rightarrow B$.
Calculating the cross-section with the usual rules \cite{BjDrell},
for a $2\rightarrow 1$ process, there occurs an unintegrated $2\pi\delta(E_f-E_i)$
where $E_f-E_i= 0$. This is interpreted as $2\pi\delta(0)\sim T$ where $T$ is the lifetime
of the final state, {\it i.e.}, the inverse width $\Gamma$.
The cross-section is then:
\bea
\label{sig}
\sigma = \frac{g^2}{2M_0\Gamma_0}\qquad 
\eea
Likewise, the field theory calculation of the width
via the allowed process $B_0\rightarrow \bar{E}e$ is:\footnote{The decay process is the time-reversed
production process, required by unitarity, and underscores the 
lack of a no-hair theorem near threshold. In fact, 
 increasingly there are  more examples
of new kinds of classical hair \cite{Coleman,Gregory}.}
\bea
\label{gam}
\Gamma_0=\frac{g^2}{8\pi} M_0
\eea
Note that $g^2/\Gamma_0 = 8\pi /M_0$ is now determined
and therefore the cross-section is
\bea
\label{sig2}
\sigma =\frac{4\pi}{M^2_0}=\frac{4}{\pi}R_0^2
\eea
This is slightly smaller than the usual presumed geometric cross-section,
 $\sim \pi R_0^2$, which owes
to the pointlike approximation. Nonetheless,
these are comparable.

To calibrate $g^2$
we require an input for $\Gamma_0$.  For small $N$
we are far from a Hawking thermal decay process,
and there are expected to be large $1/N$ corrections.
In ref.\cite{Dvali1} the  ground-state 
decay width for small $N$ is estimated to be of
order the Planck scale $M_P\sim M_0$.  We will introduce an order-unity  parameter $\eta$
and define:
\bea
\Gamma_0\sim \frac{1}{4} \eta M_0\qquad\makebox{hence,}\;\;\; g^2\sim 2\pi\eta.
\eea
Hence our crude field theory
fit to the properties of the quantum black hole 
suggests, with $\eta \sim 1$, there is reasonably
strong coupling to the fermions
with large $g^2$.  

We note that this width is considerably larger than a
computation using the Hawking temperature $T\sim M_P^2/8\pi M_0$,
which is $T\sim a/2\pi$ 
with the acceleration, $a\sim GM/R^2$, redshifted to infinity.
However, the decay is nonthermal, 
and the mini black hole decay process is happening promptly,
at extremely short distances, and on the horizon $a$, hence $T$, is infinite. Once the constituents
have escaped to a distance of a few $\sim \lambda_0$ the system is unbound,
and $T$ at infinity is irrelevant.

If we go beyond the lowest mass threshold HBH, we will
have a tower of states, each labeled by $N$. Higher $N$
states are expected to decay via coupled channel processes
such as $B_N\rightarrow B_{N-1} + X$, or a ``balding process''
as $B_N\rightarrow S_{N} + {f}_1f_2$ where $S_N$ is a Schwarzschild black hole,
and $S_N\rightarrow S_{N-1} + X$.
The exclusive process
$B_N\rightarrow {f}_1f_2$ characterized by an effective
coupling $g_N^2$ also exists.

Integrating out the HBH tower in our crude field theory 
yields an effective Nambu--Jona-Lasinio interaction that is applicable 
 below
the threshold at a scale $M\lta M_0$:
\bea
\label{L1}
{\cal{L}}_{M}&=&-\sum\limits_{N} \left(\frac{g_N^2}{M_N^2}\right)\bar{E}_L e_R \bar{e}_R E_L \nonumber \\
&\approx &- \left(\frac{g^2}{M_0^{*2}}\right)\bar{E}_L e_R \bar{e}_R E_L 
\eea
Note that width effects, $\sim iM_N\Gamma_N/2$  in the denominator,
are suppressed since we are at momenta
$p^2 \lta M_0^2$ and the width vanishes below threshold.

In principle many black holes contribute to this interaction
in any given channel.  
DG observe, however,
that the lifetime of the $N$th occupancy state
is $\sim N^{3/2}M_P^{-1}$ (see eq.(9) of \cite{Dvali1}).
For the HBH this implies $\Gamma_N\sim  (2+N)^{-3/2}M_P$ and
hence,
$g_N^2 \sim (2+N)^{-2}$  and $g_N^2/M_N^2\sim (2+N)^{-3}$. 
This therefore suggests that the sum converges quickly,
and may be reliably approximated by the ground-state term.
However, this is a large $N$ limit, and we might expect
$g_N^2 M_N\sim $ (constant) for small $N$, and hence there
may be enhancements from several states lowest
in the tower.

$M_0^{*2}$ is renormalized by fermion loop contributions extending
from $\Lambda$ down to $M$, which we treat in the block-spin
approximation with quadratic running \cite{BHL,BH}:\footnote{One does not need 
to use the ``block-spin RG'' with its
running mass. We can simply adjust the mass $M$ to it's fixed infrared value.
This still requires the critical coupling to achieve
a small physical infrared mass, canceling the UV value.}
\bea
M_0^{*2} = M_0^2 - \frac{g^2}{8\pi^2}(\Lambda^{2}-M^2).
\eea
Here 
$\Lambda\sim p_0= \sqrt{4\pi}M_P$ is the momentum space cut-off
of the theory associated with the fundamental length cut-off.
With  $g^2=2\pi\eta $ and   $M_0= {\sqrt{\pi/2}}{M_P}$  we have for the UV terms:
\bea
M_0^{*2}&=&\left(
\allowbreak \frac{\pi}{2} -\eta \right) M_{P}^{2}+O(M^2)
\eea
The critical coupling is therefore determined
\bea
\eta =\frac{\pi }{2}\qquad g_c^{ 2}=\pi^2
\eea

Note the cut-off term is $O(\hbar)$ and we are essentially arranging
a cancellation of a quantum loop against a classical mass term.
This is a common occurrence in dynamical situations.
For example, it happens with the Coleman-Weinberg potential,
the Banks-Zaks fixed point, and in any application of the Nambu-Jona-Lasinio (NJL)
model.  This, moreover, is why we can extend the
quantum momentum up to the fundamental cut-off,
while the classical bound-states we are perturbing have lower momenta.
In normal NJL the momentum cut-off $\Lambda$ is of order the
classical mass $M_0$ requiring a much larger critical coupling,
$g^2\sim 8\pi^2$.
Presently, the loop momentum cut-off extends above the scale $M_0$
of the classically bound system, so critical $g$
is somewhat smaller.
We remark that the formation
of bound-states in relativistic field theory is conceptually different than
in the case of classical quantum mechanics.\footnote{Most notably, in the broken
phase of a Nambu-Jona-Lasinio model the constituent fermion
develops a mass $m$ while the ``Higgs boson,'' which is composed
of two fermions, has a mass of $2m$. It would be wrong to conclude that
the Higgs is then unbound; the ``binding'' starts at the scale $\Lambda>>m$.}

To us,
$g$ is effective and reflects the structure of the wave-function of the black hole.
We must fine-tune $g=g_c$ to obtain
a low mass for the composite. This assumption is essentially 
a {\em scale invariance condition} imposed on the mass:
\beq
\label{var}
M^2_P \frac{d}{d M^2_P} M_0^{*2} = 0
\eeq
hence  ${g_c^2}={\pi^2}$ is determined. The scale invariance 
condition pushes
the bound-state into the infrared physics of the theory.

This may have a more concrete basis in the context
of Weyl invariant theories. In such theories the
Planck mass is dynamically generated by a spontaneous breaking of
scale symmetry, called ``inertial symmetry breaking," which does not
involve a potential but is associated with the formation
of $M_P$ during inflation \cite{Weyl}. Here there are fields that
develop VEV's, $v_i$, and the Planck mass is a function
of these $M_P(v_i)$. There
is then an exception to the statement that $\sim M_P^{-1}$ is the 
shortest distance scale, since we can deform the fields in
a Weyl invariant theory to lift $M_P$ to arbitrarily larger values.
Then, parameters of the black hole,
such as $g^2$ may be a function of ratios of these VEV's, $g^2(v_i/v_j)$.
Locally varying the VEVs may lead to
the relaxation of the black hole mass if  $d M_0^{*2}/d v_i  = 0$. This
may be interpreted as a condensate of dilatons localized around the black hole. 
At present we do not know how to implement these ideas and will
content ourselves with the fine-tuning, which is
equivalent to the usual fine tuning in the SM.

We now introduce a weak isodoublet auxiliary field $H$ that factorizes the 
interaction of eq.(\ref{L1}):
\bea
\label{L2}
{\cal{L}}_{M}=-g(\bar{E}_L e_R H + h.c.) - M_0^{*2}H^\dagger H 
\eea
Solving the equations of motion for $H$ and substituting back into
eq.(\ref{L2}) yields eq.(\ref{L1}).  This is 
our main point, that the HBH's can be virtual yet induce
a strong interaction below the scale $M_0$.  $H$ is the induced
composite scalar state due to these strong interactions from virtual HBH's.

We can now integrate the theory down to lower mass scales.
It useful to consider just the fermion loops by themselves at one-loop order,
to obtain, \cite{BHL,BH}:
\bea
\label{L3}
{\cal{L}}_{m}&=& -g(\bar{E}_L e_R H  + h.c.) - M_m^2 H^\dagger H
\nonumber \\
&& + ZDH^\dagger DH-\frac{\lambda}{2}(H^\dagger H)^2 .
\eea
Here we have displayed the induced relevant operator terms.
The ``block spin renormalization group'' keeps both
the logarithmic and
the quadratic running of the mass induced by fermion loops.

We obtain from the fermion loops \cite{BHL,BH}:
\bea
M_m^2 & = & M_0^{*2} - \frac{g^2}{8\pi^2}(M^2-m^2)
\nonumber \\
 & = &  M_0^2 - \frac{g^2}{8\pi^2}(\Lambda^2-m^2)
\eea
With critical coupling we see that $M_m^2\rightarrow 0$
with $m^2\rightarrow 0$. The running of $M_m^2$ to zero will be cut-off
by an explicit scale breaking mass term, that specifies the physical composite
Higgs doublet mass in the infrared,  $\sim 10^2$ GeV to $10^5$ TeV range.
We do not have a theory of these infrared masses at present
but fit them to the observed infrared physics.

Likewise, we have the induced wave-function renormalization constant
and the quartic coupling \cite{BHL,BH}:
\bea
Z = \frac{g^2}{16\pi^2} \ln\left( \frac{\Lambda^2}{\bar{m}^2 } \right)
\qquad
\lambda  = \frac{g^4}{8\pi^2} \ln\left(\frac{\Lambda^2}{\bar{m}^2}  \right).
\eea
The renormalized theory is then:
\bea
\label{L3}
{\cal{L}}_{m}&=& -\bar{g}(\bar{E}_L e_R H  + h.c.) - {\bar{M}}_m^2 H^\dagger H
\nonumber \\
&& + DH^\dagger DH-\frac{\bar{\lambda}}{2}(H^\dagger H)^2 .
\eea
where,
\bea
\bar{g} \sim \frac{g}{\sqrt{Z}},\;\;\;\;\; \bar{\lambda}=\frac{\lambda}{Z^2}, \;\;\;\;\;
{\bar{M}_m^2} =\frac{\bar{M}_m^2}{Z}. 
\eea
We have only used the fermion loops, which is technically justified in
a large $g^2$ limit. From this we can infer the behavior 
of the renormalized couplings as $m\rightarrow \Lambda$:
 \bea
 \label{bc3}
\bar{g}^2=\frac{1}{2}\bar{\lambda} \sim \frac{16\pi^2}{{\ln(\Lambda^2/m^2)}}\rightarrow \infty 
\eea
which is a behavior identical to the top condensation
models \cite{BHL,BH}. Note the critical coupling, $g^2=g_c^2$ cancels
in the running couplings. This corresponds to the RG running
of these couplings in the limit of retaining only the fermion loops.

Given the boundary conditions on the 
running couplings as $m\rightarrow \Lambda$
of eq.(\ref{bc3}),
we can switch to the full RG equations 
including gauge couplings, $g^2$, and $\lambda$,  {\it etc.}
To apply this to the electron we integrate the full RG equations down to a
mass scale of order $\sim 10^5$ TeV and stop. There we
install an explicit mass for the composite Higgs, $\sim M^2_{m} \sim (10^{5})^2$ TeV$^2$. This will 
then be a heavy doublet 
that does not directly develop a VEV.  However, by mass
mixing with the SM Higgs boson, $\sim \mu^2 \sim (10^{2})^2$ TeV$^2$ 
the heavy electron Higgs will acquire a tiny ``tadpole'' VEV,
$\sim v\mu^2/M_m^2 \sim 10^{-6} v$, which determines the electron mass \cite{HMTT}.

Essentially, the Higgs boson is the threshold black hole, pulled into
the far infrared by the fermion loops and the fine-tuning condition. 
The black hole is only
present at extremely short distances and is in effect virtual.
The Higgs wave-function is mainly virtual fermions and gauge fields at large
distances, triggered by the binding due to the virtual black hole 
at the Planck scale.

\subsection{Quarks}

We assume that the incident flavors
are a pair  consisting of the
top quark doublet
 $T^i_L = (t,b)_L$  with $ [I_3=(1/2,-1/2), Y= 1/3]$ 
 and right-handed singlet $\bar{t}_{j}R\sim [0, -4/3]$
 where $i,j$ are color indices. 
 
 The Lagrangian is:
\bea
DH_{j}^{i\dagger} DH_{i}^{j}-M_{0}^{2}H^{i\dagger}_{j}H_{i}^{j}-g\left( \overline{T}%
_{L}^{i}t_{jR}H_{i}^{j}+hc\right) .
\eea
Therefore the produced HBH black hole, $H_{ij}$, will be a weak isodoublet
will have  $H_{ij}\sim [I_3=(1/2, -1/2), -1] $, and its electric charge
will be $Q = [0,-1]$, identical to the SM Higgs doublet. However, it now carries mixed
color indices $i,j$ that we wish to project onto $SU(3)$ representations.

Define:
\bea
H_{i}^{j}=H^{a} \frac{(\lambda^a)^{j}_{i}}{2}
+\frac{1}{\sqrt{N_c}}H\delta _{i}^{j}
\eea
where $N_c=3$,
and we use Tr$\left( \frac{\lambda ^{a}}{2}\frac{\lambda ^{b}}{2}\right) =\frac{1%
}{2}\delta ^{ab}$ and $\left( \frac{\lambda ^{a}}{2}\right) ^{\dagger }=%
\frac{\lambda ^{a}}{2}$.
The terms in the action become:
\bea
&&
DH_{j}^{i^\dagger }DH_{i}^{j}
=\frac{1}{2}DH^{a}DH^{a} + DH^{^\dagger }DH
\nonumber \\
&&
M_{0}^{2}H^{i\dagger}_{j} H_{i}^{j} = \frac{1}{2}%
M_{0}^{2}H^{\dagger a}H^{a}+M_{0}^{2}H^{^\dagger }H
\nonumber \\
&&
g \overline{T}_{L}^{i}t_{jR}H_{i}^{j} = g\overline{T}_{L}\frac{%
\lambda ^{a}}{2}t_{R}H^{a}+g^{\prime }\overline{T}_{L}t_{R}H 
\eea
and
where $\overline{T}_{L}t_{R}=\overline{T}_{L}^{i}Ht_{jR}\delta _{i}^{j} $ 
and we have:
\bea
\frac{g}{\sqrt{N_c}}=g^{\prime }.
\eea
The decay width is now:
\bea
\Gamma =\frac{N_{c}g^{\prime 2}}{8\pi 
}M_{0}=\frac{g^{2}}{8\pi }M_{0}.
\eea
The cross-section is:
\bea
\sigma =\frac{g^{2}}{2M_{0}\Gamma }=\frac{g^{\prime 2}}{%
2M_{0}\Gamma N_{c}}.
\eea 
per color pair and 
$\frac{g^{\prime 2}}{2M_{0}\Gamma }$ color averaged.
The loop correction to the Higgs mass is as before,
\bea
{M}_m^{2}=M_{0}^{2}-\frac{%
N_{c}g^{\prime 2}}{\allowbreak 8\pi ^{2}}( \Lambda ^{2} -m^2)
\eea
and the critical coupling is $g_c^2=\pi^2=N_cg_c'^2$, hence
$0 =M_{0}^{2}-N_cg_c'^2\Lambda^2/8\pi^2$.
Likewise, we have the induced wave-function renormalization constant
and the quartic coupling:
\bea
Z = \frac{N_cg'^2}{16\pi^2} \ln\left( \frac{\Lambda'^2}{\bar{m}^2 } \right)
\qquad
\lambda  = \frac{N_cg'^4}{8\pi^2} \ln\left(\frac{\Lambda^2}{\bar{m}^2}  \right).
\eea
The renormalized parameters are:
\bea
\bar{g}' \sim \frac{g'}{\sqrt{Z}},\;\;\;\;\; \bar{\lambda}'=\frac{\lambda}{Z^2}, \;\;\;\;\;
{\bar{M}_m^2} =\frac{\bar{M}_m^2}{Z}. 
\eea
Note the quartic coupling receives a loop factor of $N_c$, not $N_c^2$. Hence
the renormalized quartic coupling will be $\sim 1/N_c$ relative to the lepton case.
This preserves the UV relation $\bar{g}'^2 \sim \bar{\lambda}'/2$.

\section{Scalar Democracy}

\subsection{Counting Higgs Black Holes}

 We can count the number of composite scalars produced by threshold
 RN-black holes.
The SM fermionic fields consist of ${48}$ two-component left-handed 
spinors, $\psi _{A}^{i}$, including all left-handed and anti-right-handed fermions.
 $ SU(48)\times U(1)  $ is then an approximate  dynamical symmetry
 (neglecting gauge interactions). 
 
 The
most general non-derivative ($s$-wave) scalar-field  bilinears 
coupled to RN-black holes takes the form:
\bea
g\epsilon^{AB}\psi_{A}^{i}\psi_{B}^{j}B_{ij}+h.c.,
\eea
where $B _{ij}$ transforms as the symmetric $ \mathbf{1176} $ representation of  $SU(48)$. 
The field $B _{ij}$ contains many complex scalar fields with assorted quantum numbers, 
including baryon and lepton number, color, and weak charges.
This describes all fermion pair collisions in the SM that can produce a black hole.

The $\bf{48}$ consists of the  $24$ 
left-handed quarks and leptons, $\Psi_{Li}$, and  $24$ right-handed 
counterparts, $\Psi_{R\widehat{i}} $.
 The index $i$ now runs over the chiral $SU(24)_{L}$ and $\widehat{i}$ 
 over the chiral $SU(24)_{R} $ subgroups of $ SU(48) $. We thus have:
\bea
\label{fields}
\Phi_{i\widehat{j}}\overline{\Psi }_{L}^{i}\Psi _{R}^{\widehat{j}}+\Omega
_{ij}\overline{\Psi }_{L}^{i}\Psi _{R}^{jC}+\widehat{\Omega }_{\widehat{ij}%
}\overline{\Psi }_{R}^{\widehat{i}}\Psi _{L}^{\widehat{j}C}+\text{h.c}.,
\eea
where $ \Phi_{i\widehat{j}} $ is the $ (\mathbf{24}_L,\, \mathbf{24}_R) $ complex scalar 
field with $24^{2}=576$ complex degrees of freedom. 
$\Omega $ and $ \widehat{\Omega } $ are the symmetric $\mathbf{300} $ representations of 
$SU(24)_{L}$ and $SU(24)_{R}$ respectively, matching  the degrees of freedom of $B _{ij}$.
Here $\Omega _{ij}$ and $\widehat{\Omega }_{ij}$ are the analogues of Majorana masses 
and carry fermion number, while $\Phi $ contains fermion number neutral fields, 
such as Higgs fields, in addition to ($B-L$) leptoquark multiplets and colored Higgs doublets.

%\newpage
The resulting spectrum of composite states in the $\Phi ^{ij}$ system becomes:
\begin{itemize}
\item $ 18 \times (\mathbf{1}, \, \mathbf{2},\, \tfrac{1}{2}) \sim \bar{Q}_{L} (U_{R}, D_{R}) $;  
Higgs doublets in quark sector $ =2\times 3^2 \times 1\times 2 = 36$ dof's)),

\item $ 18 \times (\mathbf{1}, \, \mathbf{2},\, -\tfrac{1}{2}) \sim \bar{L}_{L} (N_{R}, E_{R}) $;  
Higgs doublets in lepton sector  $ =2\times 3^2 \times 1\times 2 = 36$ dof's),

\item $ 9 \times (\mathbf{8},\, \mathbf{2},\, \pm \tfrac{1}{2}) \sim \bar{Q}_L \lambda^{a} (U_R, D_R) $; color octet, isodoublets,
$ 3^2 \times 8 \times 2 \times 2 = 288$, complex DoFs, 

\item $ 9 \times (\mathbf{3},\, \mathbf{2},\, \tfrac{1}{6} [-\tfrac{5}{6}] ) \sim \bar{L}_L  (U_R, D_R) $;
color triplet, isodoublets, $ 3^2 \times 3 \times 2 \times 2 = 108$  DoFs,

\item $ 9 \times (\bar{\mathbf{3}},\, \mathbf{2},\, -\tfrac{1}{6} [-\tfrac{7}{6}] ) \sim \bar{Q}_L (N_R, E_R) $; 
 color triplet, isodoublets, $ 3^2 \times 3 \times 2 \times 2 = 108$  DoFs,
\end{itemize}
where the brackets denote the SM quantum numbers. 
The first two entries in the above list are the $36$ Higgs doublets,
$18$ in the quark and $18$ in the lepton
sectors respectively.

The key feature is that
these bound-states will have a universal Higgs-Yukawa coupling $g$ at the scale $M_P$.
For the picture we have just outlined to work, $g$ must be sub-critical.
Otherwise,
with a supercritical coupling, $\Phi ^{ij}$ 
will condense with a diagonal VEV, $\left\langle \Phi
_{ij} \right\rangle = V \delta _{ij}$ and all the fermions would acquire large, 
diagonal constituent masses of order $gV$, grossly
inconsistent with observation.

We  assume that  $\Omega _{ij}$, $\widehat{\Omega }_{ij}$ and all color-carrying weak doublets
have very large positive $M^{2}$ and therefore we will ignore them. They will be inactive
in the RG evolution (though they may be welcome when gauge unification is included).
With $g$ taking on a nearly-but-sub-critical value for the color
singlets, the Higgs bound-states will 
generally have positive masses that can be much lighter than $M_P$.
The colored states are presumably more massive owing to the
gluon field in the RN solution
(this is a long story we'll not enter into 
presently).
Small explicit masses are introduced by hand 
as scale symmetry-breaking effects, required to split the spectroscopy
in the infrared and accommodate phenomenology.  

In scalar democracy  the  flavor physics and fermion mass hierarchy problems are 
flipped out of $d=4$ Higgs-Yukawa (HY) coupling textures and into the structure of the 
the mass matrix of the many Higgs fields.   We have no theory
of the small input masses at present, 
but we can choose these to fit the observed quark and lepton sector masses
and CKM physics, as  well as maintain consistency with constraints from rare weak decays, etc.  
It is not obvious {\em a priori} that there exists a consistent solution with the flavor 
constraints, however, it does work \cite{HMTT}.
Many of  these  mass terms are technically natural, protected by the $SU(48)$ symmetry structure
which can be seen in
a subset model in \cite{HMTT2}.
The critical theory will thus
contain many light  composite Higgs doublets with a spectrum of positive $M^2$'s
that extends from $\sim10^2$ GeV up to $\sim 10^6$ TeV. 

We refer the reader
to \cite{HMTT,HMTT2,HP} for more of the phenomenology of this ``scalar democracy,'' 
including production and detection at the LHC and upgrades.

\subsection{RG Solution}

The induced couplings $\bar{g}, \bar{g}', \bar{\lambda},\bar{\lambda}' $ satisfy
the RG for the logarithmic running below the Planck scale
(we will omit the overline in the following). 
The boundary conditions are 
determined by the binding dynamics at the Planck scale:
 \bea
 \label{bc}
({g},\;\; {g}',\;\;{\lambda},\;\;{\lambda}') \rightarrow \infty 
\eea
Presently we will only sketch very roughly the results for the $g^2(m)$ and $g'^2(m)$ RG
evolution and leave a more detailed study including the quartic couplings to \cite{HillThomsen}.

At a first glance, note that
the HY coupling of the top quark in the SM 
would be driven to the infrared-quasi-fixed point 
of  Pendleton--Ross \cite{PR} and Hill \cite{FP}. 
The skeletal RG equation
for the top quark HY coupling in the SM, $g_t$, is:
\beq
\label{one}
D g_t = g_t\left(\left(N_c+\frac{3}{2}\right)g_t^2 - \left(N_c^2-1\right)g_3^2\right)
\eeq
where $D= 16\pi^2 \partial/\partial \ln(m)$, and $m$ is
the running mass scale,  $g_3$  the QCD coupling.
For illustrative purposes we discuss the one-loop RG equation 
and suppress electroweak corrections, though they are included the figure results.

Starting the running of $g_t(m)$ at very
large mass scales, $m = M_X$, with large initial values, i.e., $g_t(M_X)>>1$
(effectively a Landau pole at $M_X$), 
it is seen that $g_t(m)$  flows into an ``infrared 
 quasi-fixed point.''
This is ``quasi'' in the sense that, 
if the QCD coupling, $g_3$, was a constant
then $g_t$ would flow to an exact conformal fixed point.  
%Scale invariance
%is therefore broken by the QCD $\beta$-function, and dictates a slow
%evolution of $g_3$, and  $g_t$ locks into the quasi-fixed point
%trajectory.
The  low energy prediction of the top quark  HY coupling 
is very insensitive to its precise, large initial
values and mass scales. Starting at $M_X=M_P$ the result comes in about $16\%$ higher than experiment.
This is shown in Fig.(1)
where the effective top mass $m_{top}=g_t(m) v $  (where $v = 175$ GeV is the electroweak scale)
is plotted vs. renormalization scale $m$; the physical top mass corresponds 
to $m\sim v$, or $\ln(m)\sim 5$.

\begin{figure}[t!]
\vspace{0.0 in}
	\vspace{-3.0 in}\hspace*{-3.6 in}\includegraphics[angle=-90,scale=1.0,width=1.5\textwidth]{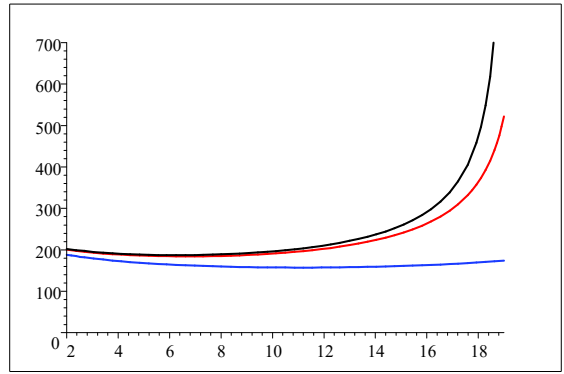}
	\vspace{-3.0 in}
	\caption{The Pendleton-Ross-Hill \cite{PR,FP} infrared-quasi--fixed point in the top Higgs-Yukawa coupling in
	the SM. Plotted is the effective top mass $m_t=g_t(m)v$ (where $v=175$ GeV) vs the log of the running scale, $\log_{10}(m/\makebox{GeV})$. The focusing in the infrared and its relative insensitivity to initial values is indicated. Initial values 
	are input at $\log_{10}(M_P/\makebox{GeV})=19$, 
	are $ g_t(M_P)=1$ (magenta); $g_t(M_P)=3$ (red); $g_t(M_P)=10$ (black).
	The SM prediction
	is about $m_t=g_t(m_t)v\approx 200$ GeV about $16\%$ above the experimental value.}
	\label{RGFlowgtopSM}
\end{figure}

However, we now expect $18$ Higgs doublets in the quark sector,
and each doublet coupled to a particular color singlet
pair, $\bar{\psi}^a_L\psi^b_R$, where $a$ counts the $3$ LH flavor doublets and
$b$ the $6 $ RH singlets. Likewise, we have
$18$ doublets in the lepton sector.
The key feature of gravitational binding
of these composite Higgs bosons is that the theory has one universal HY coupling $g'$
in the quark sector, and $g$ in the lepton sector, defined
at the Planck scale by eq.(\ref{bc}).  For the quark sector, $g'$  is determined by
the top quark HY coupling at low energies, $g'(m_t) \sim 1$.  This will be different
than the SM prediction of Fig.(1) owing to the presence of the $17$ other doublets
(as we see below).

The quark and lepton subsectors resemble $SU(6)_L\times SU(6)_R$
linear $\Sigma$-model Lagrangians, where the interaction is subcritical and ultimately
only the SM Higgs condenses. We have only observed the lightest
Higgs boson doublet thus far; the remaining doublets are massive
but mix with the SM Higgs and thus give power-law
suppressed HY couplings to the SM Higgs
hence power-law suppressed masses and mixings to the light fermions. 
The theory is predictive
and the sequential massive Higgs, $H_b$ will couple to
$\sim g'(\bar{t},\bar{b})_L H_b b_R$
(see \cite{HMTT}).

In the quark and lepton sectors, each containing $18$ doublets,
the RG equation for the universal
HY couplings take the one-loop form
\cite{HMTT}\cite{HillThomsen}:
\bea
\label{two}
D g' & = &  g;\left((3+N_f)g'^2 - (N_c^2 -1)g_3^2-\frac{9}{4}g_2^2-\frac{17}{12}g_1^2\right)
\nonumber \\
D g & =&  g\left(\left(\frac{5}{2}+N_f\right)g^2 -\frac{9}{4}g_2^2-\frac{15}{4}g_1^2\right)
\eea
where $N_f=6$ and $N_c=3$.

In Fig.(2) we show how the quark couplings $g'$ evolve
into the IR given the boundary conditions of eq.(\ref{bc}). 
This directly describes the top quark mass,$g'=g_t$, while
all other quarks (and leptons) will couple through power-law
suppressed mixing effects via their own Higgses, and
receive smaller masses \cite{HMTT}.

\begin{figure}[t!]
\vspace{0.0 in}
	\vspace{-3.0 in}\hspace*{-3.6 in}\includegraphics[angle=-90,scale=1.0,width=1.5\textwidth]{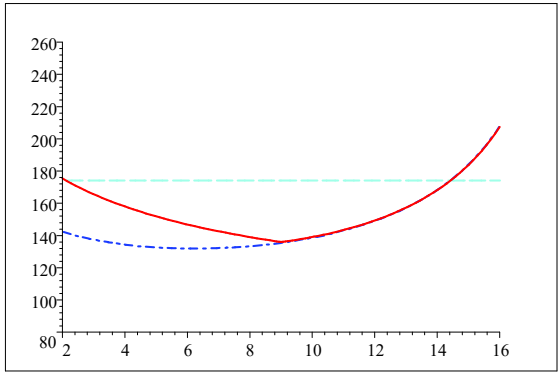}
	%{imagea.png}
	\vspace{-3.0 in}
	\caption{The running top mass plotted $175(\makebox{GeV})\times g'(m)$ (red) vs scale $\log_{10}(m/\makebox{GeV})$ in the $SU(6)\times SU(6)$ model.  We assume SM three generation running of gauge couplings, and decoupling of all but the lightest Higgs bosons at an average 
	scale of $M \sim 10^6$ TeV. This leads to the inflection point at $M$. The top mass without the decoupling is shown (dashed blue line) and the experimental top mass (magenta line).  In a more detailed scenario with improved gauge coupling running
	we expect $M\rightarrow 10^3-10^4 $TeV will be significantly lower.} 
	\label{RG4c}
\end{figure}

\begin{figure}[t!]
\vspace{0.0 in}
	\vspace{-3.0 in}\hspace*{-3.6 in}\includegraphics[angle=-90,scale=1.0,width=1.5\textwidth]{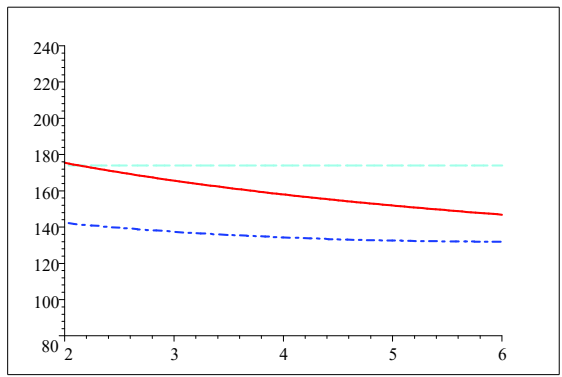}
	%{imagea.png}
	\vspace{-3.0 in}
	\caption{Detail of Fig.(2), plotting $g'(m)v$ (red) vs  $\log_{10}(m/\makebox{GeV})$ in the case of $SU(6)\times SU(6)$ with decoupling at $\sim 10^6$ TeV, while the dashed curve is the no-decoupling case. The predicted top mass, $g'(m_t)v$ for $\log_{10}(m=175)\approx 2.2$ is now in concordance 
	with experiment (magenta line) and probes the average heavy Higgs field masses in the $M \sim 10^6$ TeV range.} 
	\label{RG4d}
\end{figure}

We have a limitation in modeling since we need to know
the running of the gauge couplings  $g^2_i$, given the large multiplicity
of Higgs fields. This also poses challenges for gauge unification. 
Preliminary studies indicate that gauge unification is possible, but it
is likely to be more complex than the usual picture.
Here we 
simply use the SM values for the $g_i^2$, while with the large multi-Higgs spectrum
we expect larger values.

Naively applying the $N_f=6$ with SM $g_i^2$ it 
appears in Fig.(2) that the top mass
undershoots the experimental result (blue dashed curve).
However, the effect of decoupling of the heavier
Higgs bosons causes the $g' $ to come up to 
concordance with the $174$ GeV observed
value (red curce).  The observed top mass is therefore sensitive
to the extended Higgs sector decoupling scale, $M$, (just as the top and $W$ masses
were sensitive to the Higgs boson and predicted its discovery mass).

As stated above, in scalar democracy we explain the origin of mass 
and CKM mixing in the SM in a novel way---all flavor physics is mapped into the masses and 
mixings of the array of composite Higgs bosons which have universal couplings to their particular
constituent fermion bilinears. The lowest eigenmode is the SM Higgs boson,
corresponding to $\bar{t}t$.

It is somewhat easier to grasp the details of the theory by
focusing on the $t$-$b$ subsector as in \cite{HMTT2}.
Here we  predict the first
sequential Higgs $H_b$ with the large $g'\sim O(1)$ coupling
to $g'\bar{T}_LH^c_bb_R$  (with some additional QCD RG  flow
to the b-quark mass from $5$ TeV 
this coupling is $g'\sim 1.5$). We expect  an upper mass bound 
on $H_b$ of order $5.5$ TeV.
This state is 
accessible at the LHC or its upgrade (see
\cite{HMTT} and for the third generation predictions, \cite{HMTT2})). 
Above all, we predict the key result that sequential Higgs bosons couple 
with a common (modulo renormalization group effects) $O(1)$ coupling, 
as calibrated by the top quark Higgs-Yukawa coupling constant 
and dictated by the RG infrared quasi-fixed point.  The observation of
the $H_b$ with $g\sim g_{top}$ would offer significant support to this scenario.

\section{Conclusions}

We are thus led to a 
new idea:  Higgs bosons are composite bound-states of standard model 
fermion pairs driven by threshold black holes at $M_P$ with
the corresponding quantum numbers. The black holes of the far UV  are quantum
mechanical, mini black holes that are dressed
by fermion loops to acquire lower energy (multi-TeV scale) masses.
There are many bound-state Higgs bosons, at least one per
fermion pair at $M_{Planck}$, and a rich spectroscopy of Higgs bosons is expected to emerge.
This theory dynamically unifies Planck scale physics with the electroweak
and multi-TeV scales.  By studying Higgs physics at the LHC one may have a window
on the threshold spectrum of black holes at the Planck scale.

The production and decay of the these states has
been modeled by effective field theory vertices and masses, and leads
to a Nambu-Jona-Lasinio effective field theory for the composite Higgs bosons.
We estimate of the critical coupling $g_c^2\sim \pi^2$ (or $g_c'^2 \sim \pi^2/3$).
In the NJL model there is fine tuning of $g=g_c$, at the same level as
occurs in the SM. This can be stated as 
a scale invariance condition imposed on the composite mass at
the Planck scale.  We would hope to someday replace
the fine-tuning by a dynamical phenomenon, perhaps involving an
underlying scale invariance of the full theory.

This scenario provides an underlying 
dynamics for the recently proposed ``scalar democracy'' \cite{HMTT,HMTT2},
in which every fermion pair in the SM 
in an $s$-wave combination is argued to be associated with a gravitationally
bound composite Higgs field. 
A consequence of this hypothesis is that the many resulting Higgs
bosons couple universally to matter. This can explain the
masses and mixings of fermions, not by textures,
but rather via the masses and mixings
of the many Higgs doublets. 

Here the observed SM Higgs isodoublet is a $H \sim \bar{t}_R\times (t,b)_L$
composite. The HY coupling of the top
quark calibrates the universal coupling of all Higgs bosons, modulo RG effects.
The Higgs-Yukawa universality is
a critical prediction of the scenario and its gravitational underpinnings. It can be tested by
finding at the LHC (upgrade) the first sequential heavy Higgs doublet,
the $H_b$ with a mass of $\lta 5.5$ TeV, and confirming
its HY coupling to $\bar{b}b$  is $O(1)$ \cite{HMTT2,HP}.
We view the search for sequential Higgs doublets with $O(1)$
HY couplings at LHC to be of high importance.
These states will occur
in any given channel defined by any SM fermion pair
owing to the universality of gravity.

This dynamical picture we've presented depends upon the properties
of mini black holes in the quantum limit and strong coupling.
We have followed a simple
schematic model of quantum black holes due to Dvali and Gomez,  \cite{Dvali1,Dvali2,Dvali3},
which we find compelling, and similar to the Bohr model of the Hydrogen atom.  
The DG model is a kind of ``bag model'' of quantum black holes 
with a fundamental length cut-off.
We extend the model to include fermion pairs.

While it would seem that we are introducing many new scalars, 
in fact, we are reducing the number of fundamental degrees of freedom
of the SM by replacing the existing Higgs by a gravitational bound-state.
The rich dynamics of gravity is argued to produce the complexity of multiple
scalar fields the infrared.   Perhaps this can be extended to explain the
origin of flavor and, more radically, composite gauge fields. We will
return to these questions elsewhere.

A spectrum of many Higgs scalars with $O(1)$ Yukawa couplings
would dramatically contradict the present day paradigm of weak coupling, simple unification, 
and string theory.   
The Higgs sector is usually constrained by theorists to a few doublets.  
We believe in the future this will change,  perhaps 
with LHC ugrades or beyond, where the Higgs
sector may spring to life.  Scalar democracy should be
taken seriously as a reasonable hypothesis for this as it can
explain the CKM mixing and mass hierarchies seen in the matter sector of the SM.
It is intrinsically a composite theory of scalars. 
If a few of its states are
confirmed we believe this will indicate a deeper gravitational, non-string
theoretic connection such as
we have described here.

\newpage

\section*{Acknowledgments}

We thank W. Bardeen and G. G. Ross for discussions
and
the  Fermi Research Alliance, LLC under Contract No.~DE-AC02-07CH11359 
with the U.S.~Department of Energy, 
Office of Science, Office of High Energy Physics.

\end{document}